\documentclass[aps,
12pt,
final,
notitlepage,
oneside,
onecolumn,
nobibnotes,nofootinbib,superscriptaddress,noshowpacs]
{revtex4-2}

\usepackage{amsmath,amssymb}
\usepackage[utf8]{inputenc}
\usepackage[english]{babel}
\usepackage{hyperref}
\usepackage{caption2}
\usepackage{graphicx}

\begin{document}


\title{Parallel Version of CORSIKA Code with Cherenkov Option for SPHERE-3 Project}

\author{\firstname{M.~D.}~\surname{Ziva}}
\email[E-mail: ]{maxim.ziva@gmail.com}
\affiliation{Skobeltsyn Istitute for Nuclear Physics, Lomonosov Moscow State University, Lenenkie gory, 1(2), Moscow, 119234, Russia}
\affiliation{Faculty of Computational Mathematics and Cybernetics, Lomonosov Moscow State University, Leninskie gory, 1(52), Moscow, 119234, Russia}

\author{\firstname{V.I.}~\surname{Galkin}}
\email[E-mail: ]{v_i_galkin@mail.ru}
\affiliation{Skobeltsyn Istitute for Nuclear Physics, Lomonosov Moscow State University, Leninskie gory, 1(2), Moscow, 119234, Russia}
\affiliation{Faculty of Physics, Lomonosov Moscow State University, Leninskie gory, 1(2), Moscow, 119234, Russia}

\author{\firstname{E.A.}~\surname{Bonvech}}
\email[E-mail: ]{bonvech@yandex.ru}
\affiliation{Skobeltsyn Istitute for Nuclear Physics, Lomonosov Moscow State University, Leninskie gory, 1(2), Moscow, 119234, Russia}

\author{\firstname{O.V.}~\surname{Cherkesova}}
\affiliation{Skobeltsyn Istitute for Nuclear Physics, Lomonosov Moscow State University, Leninskie gory, 1(2), Moscow, 119234, Russia}
\affiliation{Department of Cosmic Research, Lomonosov Moscow State University, Leninskie gory, 1(52), Moscow, 119234, Russia}

\author{\firstname{D.V.}~\surname{Chernov}}
\affiliation{Skobeltsyn Istitute for Nuclear Physics, Lomonosov Moscow State University, Leninskie gory, 1(2), Moscow, 119234, Russia}

\author{\firstname{V.A.}~\surname{Ivanov}}
\affiliation{Skobeltsyn Istitute for Nuclear Physics, Lomonosov Moscow State University, Leninskie gory, 1(2), Moscow, 119234, Russia}
\affiliation{Faculty of Physics, Lomonosov Moscow State University, Leninskie gory, 1(2), Moscow, 119234, Russia}

\author{\firstname{T.A.}~\surname{Kolodkin}}
\affiliation{Skobeltsyn Istitute for Nuclear Physics, Lomonosov Moscow State University, Leninskie gory, 1(2), Moscow, 119234, Russia}
\affiliation{Faculty of Physics, Lomonosov Moscow State University, Leninskie gory, 1(2), Moscow, 119234, Russia}

\author{\firstname{N.O.}~\surname{Ovcharenko}}
\affiliation{Skobeltsyn Istitute for Nuclear Physics, Lomonosov Moscow State University, Leninskie gory, 1(2), Moscow, 119234, Russia}
\affiliation{Faculty of Physics, Lomonosov Moscow State University, Leninskie gory, 1(2), Moscow, 119234, Russia}

\author{\firstname{D.A.}~\surname{Podgrudkov}}
\affiliation{Skobeltsyn Istitute for Nuclear Physics, Lomonosov Moscow State University, Leninskie gory, 1(2), Moscow, 119234, Russia}
\affiliation{Faculty of Physics, Lomonosov Moscow State University, Leninskie gory, 1(2), Moscow, 119234, Russia}

\author{\firstname{T.M.}~\surname{Roganova}}
\affiliation{Skobeltsyn Istitute for Nuclear Physics, Lomonosov Moscow State University, Leninskie gory, 1(2), Moscow, 119234, Russia}




\begin{abstract} 
We use Lomonosov-2 supercomputing facility for the generation of extensive air shower events with Cherenkov light which is a rather time consuming procedure. At primary energies slightly below 100~PeV  a substantial part of events are killed before reaching their end while exceeding the queue time limit. The fact compelled us to develop a multithread version of the code. We report here the main features of our development as well as some evidence of its efficiency.

\end{abstract}


\keywords{cosmic rays, extensive air showers, Cherenkov light, Monte Carlo simulation,
parallel calculations, SPHERE-3} 

\maketitle

\section{Introduction}

Among the problems of cosmic ray astrophysics there are three classical ones: the energy spectrum of primary cosmic rays (PCR), their arrival direction distribution (usually called the anisotropy problem) and the assortment of arriving particles, i.e. their chemical/mass composition.
    
The problems are at least 70 year old but still far from being solved. As the known energy spectrum of PCR spans more than 12 orders, the methods used to detect them vary with the energy. The so called direct methods are applicable below $10^{15}$~eV and use satellite or high altitude balloon detectors of relatively small aperture because the PCR flux is still sufficient for the measurements. At higher energies one has to use the indirect method of extensive air showers (EAS) initiated by PCR because their flux decreases rapidly with energy.
    
EAS is a large particle cascade beginning with the first interaction of a primary particle with an air nucleus (mostly N, O or Ar). The majority of PCR are atomic nuclei, also electrons, $\gamma$-quanta and neutrinos are present among them. Primary nucleus of super high energy undergoes a hadron interaction resulting in a bunch of secondary hadrons subject to further hadron interactions with air nuclei. Multiplying hadrons form an EAS hadron skeleton. Some short-lived hadrons decay into gammas, electrons, muons and neutrinos and thus form other components of EAS. Charged particles of EAS radiate Cherenkov and fluorescent light and radio waves which may be called tertiary EAS components.
    
EAS method of super high energy PCR registration incorporates quite a number of techniques with a variety of detectors used. We are now developing a variant of Cherenkov technique suggested by A.E. Chudakov~\cite{Chudakov1972} using EAS Cherenkov light reflected from the snowed surface and observed by an air-borne telescope. The method~\cite{related} has certain advantages over the related techniques, cheapness and mobility among them. We are particularly interested in the PCR mass composition problem which presents the strongest challenge of the three problems mentioned: it requires accurate measurement of each EAS event for the assessment of the primary particle mass~\cite{mass}.
    
To optimize the method and construct a new detector of SPHERE series, SPHERE-3, we carry out vast simulations of EAS in $10^{15}$--$10^{17}$ eV energy range using Lomonosov-2 facility~\cite{SC} with CORSIKA-7 code~\cite{CORSIKA-7}. Original version of the code does not support parallel processing of the subcascades with Cherenkov light generation, thus we have to use single core for each event simulation. At energies above 70~PeV the simulation time of an event usually exceeds the limit of Lomonosov-2 job processing queue which forces us to develop a parallel version of the code.

\section{Initial modification of CORSIKA}

From the very beginning of our project, we could not use the original CORSIKA code for at least two reasons: a) our technique was not going to use particle detectors and we did not need standard CORSIKA secondary particle output files, b) we intended to create a vast artificial EAS event database with Cherenkov light characteristics according to the project requirements for further multiple use. Inevitably, we had to modify CORSIKA output to meet our needs. Besides, we also needed some additional information which is not provided by the standard version.
    
After a pertinent modification of the code a typical artificial EAS event includes: a spatial-temporal Cherenkov light  distribution at the snow level (Lake Baikal), Cherenkov light  distributions in space, angles and arrival times at three altitudes above the snow (500, 1000 and 1500~m) in histograms with binning matched to the detector characteristics. This modification made the event files significantly more compact without the loss of essential physical information about the EAS. A single event is represented by a binary file of approximately 6~GB, which compresses to less than 1~GB. The database currently occupies about 100~TB of disk space and contains more than $10^5$ events.

\section{Multithreaded version in outline}

\begin{figure}[h]
    \setcaptionmargin{5mm}
    \onelinecaptionstrue  
    \includegraphics{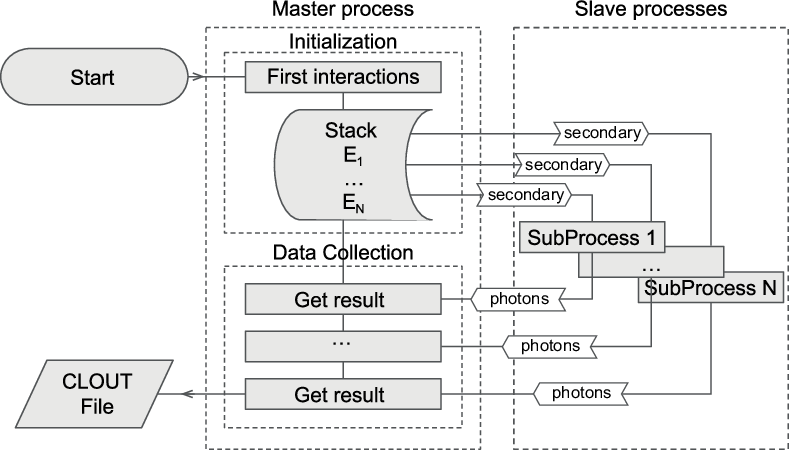}
    \captionstyle{normal}\caption{Parallel CORSIKA operation scheme}\label{fig:scheme}
\end{figure}

The main idea of our approach to transform the modified CORSIKA code into a multithreaded version was to hit two targets: (a) to distribute the computational load of tracking of a huge particle cascade with Cherenkov photon emission across multiple CPU cores and (b) to allocate only once quite a piece of memory for the multivariate arrays. We divide the process of generating an event into two stages (see outline of the parallel program operation in Fig.~\ref{fig:scheme}). At the first stage we perform the standard CORSIKA initializations and save these ready-to-use data to a binary file.

The second stage begins with reading the file. Subsequently, master thread starts to track the primary particle through the atmosphere, building up the secondary particle stack. Here we modify the standard CORSIKA random way of filling the stack so as to make the leader (the most energetic particle among the secondary particles born in an interaction) to be the last to enter the stack and thus the first to leave it. Such order of stack formation aims at tracking the leader by the master thread first. This operation is relatively fast and generates a large pool of secondary particles that can then be evenly distributed among the slave threads.

The number of leader interactions to be followed by the master thread depends on the mass of the projectile nucleus because nucleus-nucleus interactions often happen to be peripheral with low multiplicity of secondary particles and rather small loss of the projectile energy. 

The tracking of the leader stops when its energy drops below a certain threshold — for example, to approximately 2\% of the primary energy. This reduction in energy helps ensure a more balanced distribution of energy across parallel processes. While the exact threshold may vary, the goal is to reduce the leader's energy to a level that guarantees a sufficient number of secondary particles have been generated to initiate efficient parallel tracking.

At this point, the master thread divides the stack between the slave threads and starts aggregating Cherenkov photons produced in sub-cascades into multidimensional arrays. CORSIKA only generates and tracks Cherenkov photons producing photoelectrons in photosensor with unit probability.
After all slaves complete their tasks, the arrays are stored to the resulting binary file. Such distribution of duties between master and slave threads has proven to be highly effective.

\section{Stack distribution algorithm}

The process of the particle stack distribution among slave threads is crucial for accelerating event generation. When the leader tracking terminates, the master thread calculates the total energy content of the stack $E_{\text{tot}}$ and sorts the particles in ascending order of energy. The target energy per slave process is calculated as:
\[
e_p = \frac{E_{\text{tot}}}{n_{\text{slaves}}},
\]
where $n_{\text{slaves}}$ denotes the number of available computational processes.

The master thread then sequentially constructs sub-stacks for each slave. Particles are selected from the sorted list starting with the least energetic ones, and their indices are placed into a sub-stack array. The cumulative energy $e_{\text{pp}}$ of the current sub-stack is accumulated. Formation of the sub-stack for the current process terminates once $e_{\text{pp}}$ exceeds $e_{\text{p}}$. This approach ensures that the energy assigned to each process remains close to the average value, and the resulting sub-stacks contain particles with comparable total energies.  

However, in practice, uniform distribution is not guaranteed: since individual particles are indivisible, the cumulative energy in a sub-stack may either exceed $e_{\text{pp}}$ or fall significantly below it. This issue is particularly pronounced when high-energy gamma quanta are present in the stack, as their energy can occasionally constitute up to 50\% of the total energy. In such cases, a sub-stack assigned to one process may accumulate energy substantially exceeding the target value, leaving insufficient energy for subsequent processes. Consequently, situations may arise where some processes receive no particles at all.

This straightforward approach to load distribution clearly requires optimization. A simple improvement is to calculate the target sub-stack energy as $E_{\text{tot}}/(n_{\text{slaves}} + 1)$ rather than ${E_{\text{tot}}}/{n_{\text{slaves}}}$. Another consideration is that the number of slave threads should be adjusted according to the leader particle energy at the moment of stack partitioning; for instance, when the leader's energy is reduced to 2\% of the primary energy, using 10 slave threads is recommended for balanced load distribution.

Leader tracking terminates either when the leader's energy drops to 2\% of the primary energy or when a high-energy gamma quantum appears in the stack. The latter condition is motivated by the fact that gamma-initiated electromagnetic cascades are computationally expensive. Continuing sequential processing in the master thread becomes inefficient in such cases; it is more advantageous to distribute the stack among slave threads and initiate the parallel computation phase. The 2\% energy threshold was empirically selected to improve the granularity of energy partitioning among computational processes, minimizing the time spent on sequential tracking while enabling a more balanced distribution of the total energy.

Future improvements will focus on optimizing the stack partitioning algorithm to achieve more uniform particle distribution among slave processes.


\section{Testing Results}

\subsection{Test platform configuration and experimental parameters}
\label{subsec:test_platform}

For debugging the parallel algorithm and conducting preliminary tests, a local server with the following configuration was employed:
one 16-core AMD Ryzen 9 5950X processor, base frequency 3.4~GHz;
Motherboard ASUS TUF GAMING X570-PRO, AM4 socket, AMD X570 chipset, supporting up to 32 execution threads;
RAM 128~GB DDR4 ECC (4~modules of 32~GB each, frequency 3200~MHz);
Operating system Linux Fedora 38.

The server configuration provides high single-thread performance and sufficient memory capacity for simultaneous processing of multiple EAS events without disk swapping. The generation of a single event consumes approximately 7~GB of memory for the master process and about 450~MB for each slave process. Consequently, accounting for system reserves, the server can simultaneously execute up to 15 events in the sequential program version or two events in the parallel version using 10 additional processes.

This server was used for the development and testing of the parallel program version, as well as for comparing its results with those of the original version. The comparison was performed based on simulations of test event sets with the following parameters:
\begin{itemize}
\item primary particle type: proton and iron nucleus ($^{56}$Fe);
\item primary particle energies: $10^{15}$, $10^{16}$ and $10^{17}$~eV;
\item zenith angle: theta = $0^\circ$;
\item azimuthal angle: uniform distribution in the range 0--360$^\circ$;
\item hadronic interaction model: QGSJET-II~\cite{qgsIIa, qgsIIb};
\item atmospheric model: standard US atmosphere (number~1 in the CORSIKA package).
\end{itemize}

For the sequential program version, 51 events were simulated for each parameter set. For the parallel version, 11 events were simulated with the number of worker processes $n_{\text{slaves}}$=10.

\subsection{Experimental results: performance and physical validation}
\label{subsec:experimental_results}

\begin{figure}[bh]
    \setcaptionmargin{5mm}
    \onelinecaptionstrue  
    \includegraphics[width=10cm]{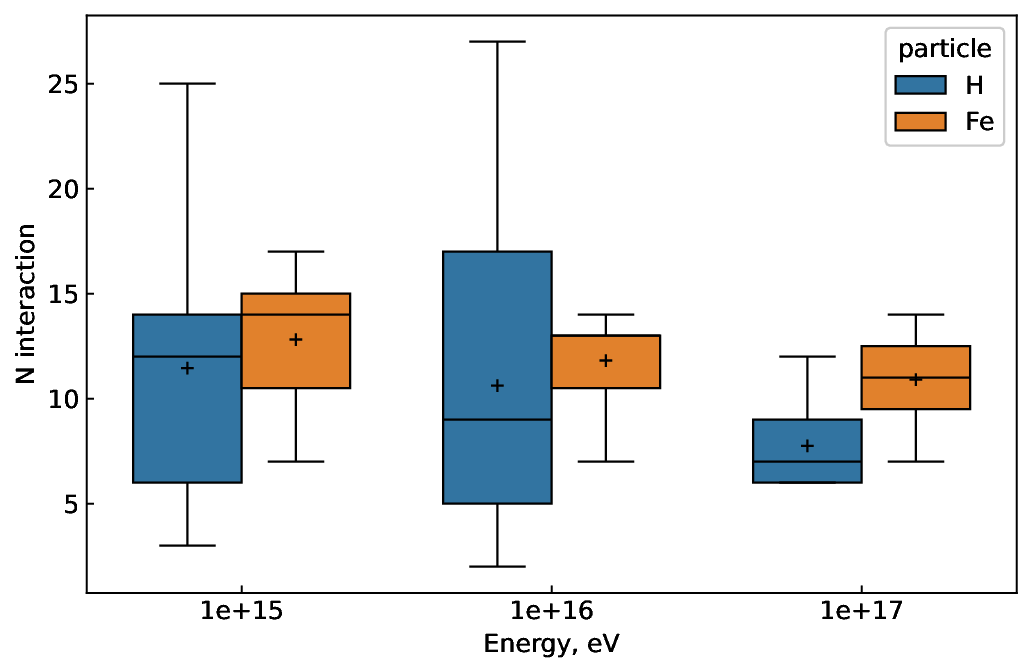}
    \captionstyle{normal}\caption{Number of primary particle interactions prior to reaching the 2\% energy threshold.}\label{fig:interactions}
\end{figure}

The statistics of primary particle interactions prior to reaching the 2\% energy threshold are presented in Fig.~\ref{fig:interactions}. Iron nuclei exhibit a larger mean number of interactions compared to protons of the same energy (e.g., 12.82 versus 11.45 at 1~PeV). The spread of values (min--max) for protons significantly exceeds that for iron nuclei. This difference is most pronounced at 10~PeV: protons show a range of 3--27 (factor of 9), whereas iron nuclei range from 7 to 14 (factor of 2). Such variability stems from fluctuations in the depth of the first interaction and the stochastic nature of nuclear cascade development, which is particularly pronounced for light primary particles. At higher energies, the multiplicity of secondary particle production per interaction increases, leading to a reduction in the total number of interactions required to dissipate the primary energy.

\begin{figure}[h]
    \setcaptionmargin{5mm}
    \onelinecaptionstrue  
    \includegraphics[width=10cm]{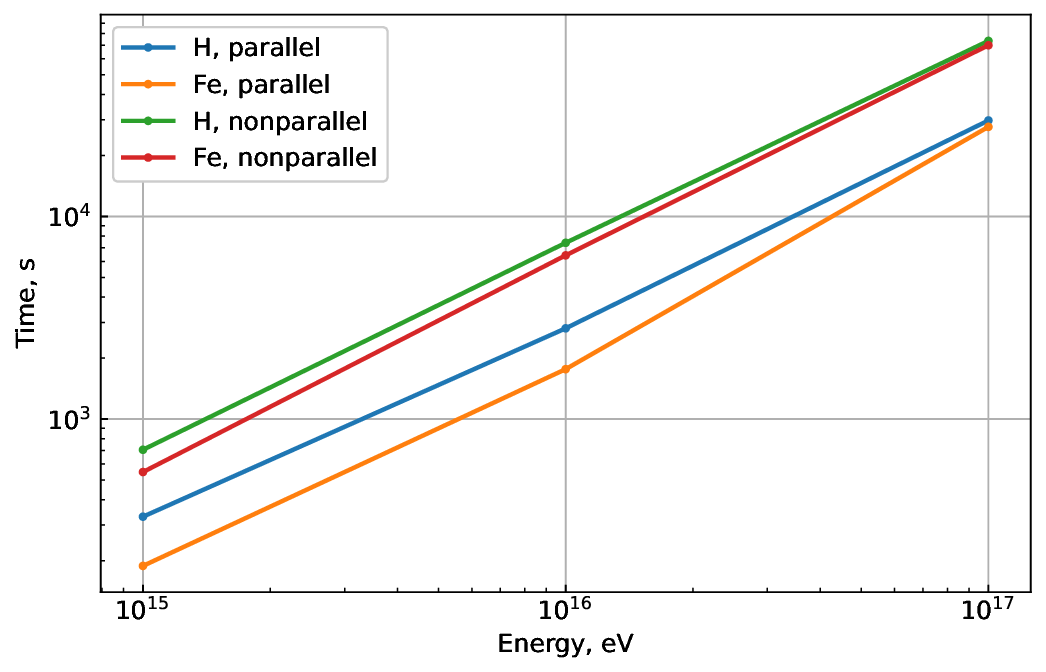}
    \captionstyle{normal}\caption{Average calculation time for one event in two versions of the program on a local server.}\label{fig:mean_time_server}
\end{figure}

\begin{figure}[h]
    \setcaptionmargin{5mm}
    \onelinecaptionstrue  
    \includegraphics[width=10cm]{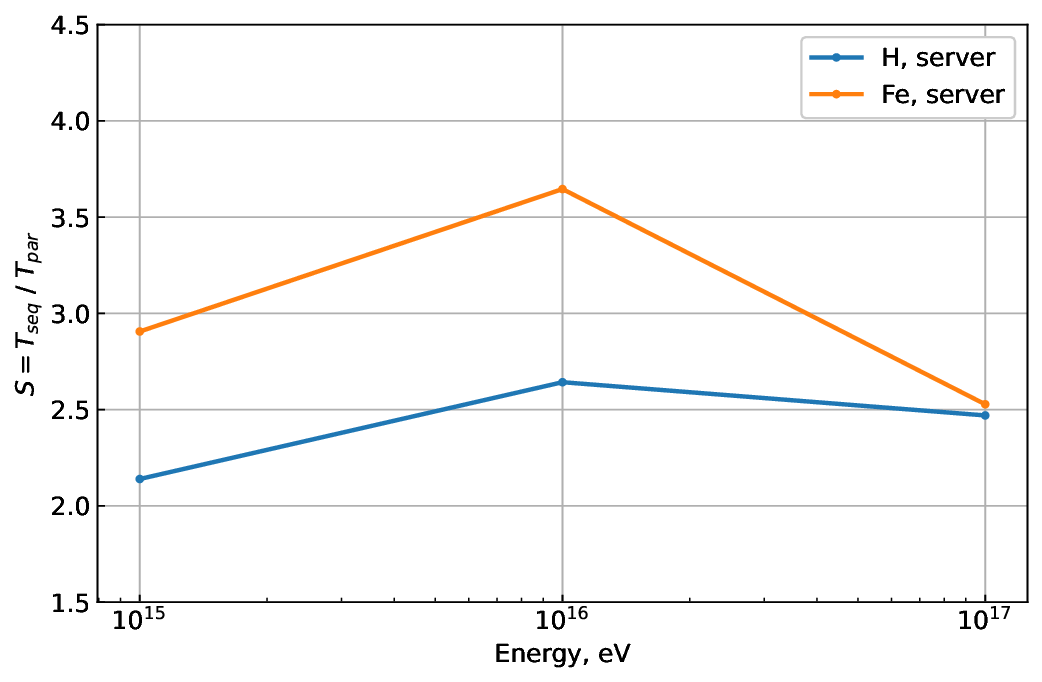}
    \captionstyle{normal}\caption{Average speedup for calculating one event}\label{fig:mean_time_server_relative}
\end{figure}

As a performance metric, we measured the execution time on the test server (see Fig.~\ref{fig:mean_time_server}). For events initiated by primary protons with energy $10^{17}$~eV, the average processing time per event decreased from 20~hours in the sequential version to 7.5~hours in the parallel version. The resulting speedup $S = T_{\mathrm{seq}} / T_{\mathrm{par}}$  for this parameter set equals 2.7. Speedup values for other primary particle types and energies are shown in Fig.~\ref{fig:mean_time_server_relative} and range from 2.2 to 3.6 depending on the energy and nuclear composition.

\begin{figure}[h]
    \setcaptionmargin{5mm}
    \onelinecaptionsfalse 
    \includegraphics[width=10cm]{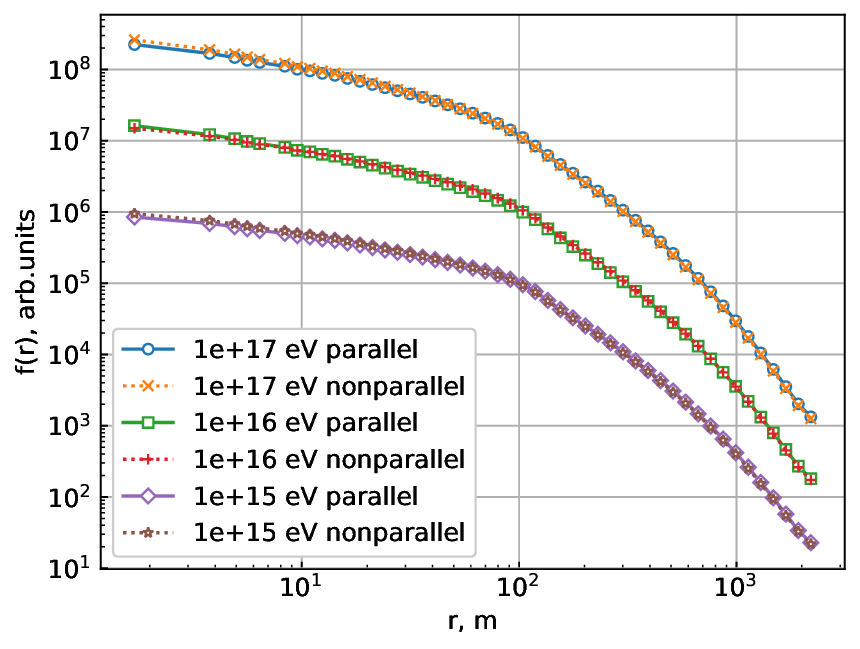}
    \captionstyle{normal}\caption{Comparison of the average spatial distribution functions of Cherenkov photons from a primary proton in two versions of the program.} \label{fig:mean_ldf_proton_server}
\end{figure}

To validate the physical correctness of the parallel implementation, we compared spatial distributions of Cherenkov photons at the observation level. Lateral distribution functions (LDFs), averaged over event statistics and obtained for identical sets of primary particle parameters, were analysed. Figure~\ref{fig:mean_ldf_proton_server} presents the averaged transverse photon distributions for primary protons of three energies ($10^{15}$, $10^{16}$ and $10^{17}$~eV). The differences between sequential and parallel versions are negligible and do not exceed the level of statistical fluctuations expected for finite sample sizes. No systematic bias is observed.

We also compared the mean total number of Cherenkov photons per event. For iron nuclei, the relative discrepancy remained within 1--4\% across the entire energy range. For protons, the discrepancy was slightly higher, ranging from 1\% to 8\%. This difference is explained by two factors. First, proton-induced showers in the considered energy range exhibit higher intrinsic variability: the relative dispersion of photon yield is substantial due to fluctuations in the depth of the first interaction and the development of the electromagnetic cascade component. Second, the difference in statistical sample sizes between the two versions leads to sampling fluctuations in the estimated means. To verify the absence of systematic errors introduced by the parallelisation algorithm, an additional experiment with identical statistical volumes is planned.

The obtained results confirm that the applied parallelisation strategy preserves the statistical properties of the generated ensembles and does not introduce systematic distortions into physically relevant distributions.

\section{Conclusions}

We have developed a multithreaded parallel version of the CORSIKA code with Cherenkov light option to overcome computational limitations encountered when simulating extensive air showers at ultrahigh energies ($10^{15}$--$10^{17}$~eV) on the Lomonosov-2 supercomputer. The implementation employs a two-stage algorithm: sequential tracking of the primary particle and its most energetic descendant (the leader) until its energy drops below 2\% of the primary energy, followed by parallel processing of the resulting particle stack distributed among slave threads.

A modification of the CORSIKA code we once conceived for the solution of our technical problem with Lomonosov-2 is now fully operable. Performance tests demonstrate a speedup factor of $S = 2.2$--$3.6$ depending on primary particle type and energy, reducing the processing time for $10^{17}$~eV proton events from 20~hours to 7.5~hours on the testing server platform. The acceleration achieved is no record but fits our expectations. 

Physical validation confirms statistical equivalence between sequential and parallel versions: lateral distribution functions of Cherenkov photons and mean photon yields agree within 1--8\% for protons and 1--4\% for iron nuclei, consistent with expected sampling fluctuations.

The main limitation of the current implementation is non-uniform particle distribution among slave processes due to the nature of high-energy particles interactions, occasionally leaving some threads idle. Future work will focus on optimizing the stack partitioning algorithm and exploring GPU acceleration in our calculations at least for some fragments of the code. Despite these limitations, the developed code already enables efficient generation of large EAS event databases for the SPHERE-3 project and can be useful for other cosmic ray experiments requiring Cherenkov light simulations.

\section{Acknoweledgements}

\begin{acknowledgments}
The study was conducted under the state assignment of Lomonosov Moscow State University, work is supported by the Russian Science Foundation under Grant No. 23-72-00006, https://rscf.ru/project/23-72-00006/, the work of V.A. Ivanov was supported by Theoretical Physics and Mathematics Advancement Foundation ``BASIS'' (grant \#24-2-10-53-1).

The research is carried out using the equipment of the shared research facilities of HPC computing resources at Lomonosov Moscow State University~\cite{SC}.

\end{acknowledgments}

%
%

\end{document}